\documentclass[prb,showpacs,amsmath,amssymb,preprint]{revtex4}

\usepackage{amsmath,graphicx}
\usepackage{epsfig}

\begin{document}

\title{Localization of a multiblock copolymer at a selective interface:\\
Scaling predictions and Monte Carlo verification} 

\author{Andrea Corsi$^{1}$, Andrey Milchev$^{1,2}$,Vakhtang G. Rostiashvili$^{1}$, 
and Thomas A.  Vilgis$^{1}$}\email{vilgis@mpip-mainz.mpg.de}
\affiliation{$^1$ Max Planck Institute for Polymer Research, Ackermannweg 10, 
55128 Mainz, Germany.\\
$^2$ Institute for Physical Chemistry, Bulgarian Academy of Science, 1113 Sofia, Bulgaria}

\begin{abstract}

We investigate the localization of a hydrophobic - polar (HP) - regular copolymer 
at a selective solvent-solvent interface with emphasis on the impact of block length
$M$ on the copolymer behavior. The considerations are based on
simple scaling  arguments and use the mapping of the problem onto a 
homopolymer adsorption problem. The resulting scaling relations treat
the gyration radius of the copolymer chain perpendicular and parallel 
to the interface in terms of chain length $N$ and block size $M$,  as well 
as the selectivity parameter $\chi$  . The scaling relations differ for 
the case of weak and strong localization. In the strong localization 
limit a scaling relation for the lateral diffusion coefficient $D_{||}$ 
is also derived.  We implement a dynamic off-lattice Monte - Carlo model to verify 
these  scaling predictions. For chain lengths in a wide range ($32 \le N \le 512$)
we find good agreement with the scaling predictions.

\end{abstract}
\maketitle

\section{Introduction}
The behavior of hydrophobic - polar copolymers (or HP - copolymer) at a selective 
interface (the interface which divides two immiscible liquids, say, water and oil,
each of them being
good for one type of monomers and bad for the other) is of great importance 
in the chemical physics of polymers. HP - copolymers are readily localized at such 
an interface because under a sufficiently large degree of selectivity the hydrophobic and 
polar parts of a copolymer chain try to stay on different sides of the interface due to 
the interplay between the entropy loss in the vicinity of the interface and the energy gain
in the proper solvent. 
This reduces the interfacial tension between the immiscible liquids and motivates the 
main practical use of such copolymers as compatibilizers \cite{Roe}. Early in the 
study of compatibilizers, the attention has been focused mainly on diblock copolymers 
due to their relatively simple structural characteristics \cite{Nool,Leibler}. These 
studies of diblock compatibilizers have been extended later by making use of the numerical 
self - consistent field approach and of Monte Carlo (MC) simulations 
\cite{Lyats1,Lyats2,Lyats3,Israels,Werner}. 
The seminal paper by Garel et. al. \cite{Garel} has attracted attention to the interface 
localization of {\em random} HP - copolymer and has brought about a great number of 
publications on this subject 
\cite{Donley,Milner,Sommer1,Sommer2,Sommer3,Sommer4,Chen1,Chen2,Denesyuk,Step}.
As a rule, the macromolecules considered in these studies pertain to random or strictly 
alternating
copolymers whereas the impact of block size on the localization transition and on the 
properties of the interface have found comparatively little attention. In fact, to the 
best of our knowledge the only MC - simulation of multiblock HP - copolymers at a 
selective interface has been carried out by Balasz et al. \cite{Balazs} within a 
two-dimensional model where no treatment in terms of scaling has been suggested.

In this paper we focus on the interesting special case of multiblock copolymers 
made of a regular sequence of HP - diblocks where  the fractions of H - and P - 
segments are equal. The main objective is to bring together simple scaling arguments 
and the findings of MC - simulations. The principal idea of the scaling consideration 
in this case implies (see ref.\cite{Sommer5,Sommer6,Leclerc}) that the problem at hand 
can be mapped onto the homopolymer adsorption problem whereby each HP - diblock (e.g.
each part of consecutive blocks made of H- and P-monomers, respectively) acts as an 
{\em effective} segment. At the interface these HP - diblocks behave as ``soft dipoles'' so 
that their ``polarization'' minimizes the free energy. This phenomenon governs the HP - diblock 
effective attraction energy, and enables  the mapping onto the problem of a homopolymer 
adsorption at a penetrable surface \cite{Gennes,Eisen}.

In Sec.II we provide refined scaling considerations which are partially based on previous 
results \cite{Sommer3,Sommer5,Leclerc}. The results are presented in such a form that the 
chain and block length dependences are clearly separated and can be easily compared with the MC 
- findings. In Sec. III we report on the results of a comprehensive MC - simulation study
and compare these results with the scaling predictions. We conclude with a brief summary
and an outlook for further work.

\section{Scaling}

\subsection{Static}

As mentioned before, the scaling treatment is based on the idea that a regular 
multiblock copolymer can be considered as a homopolymer where a single HP - diblock 
plays the role of an effective segment. All such diblocks try to keep their H - and P - 
segments in the corresponding preferable environment. This leads to the diblock  
``polarization'' at the interface and to a free energy gain which produces an 
effective attraction energy $E$. We restrict ourselves here to the fully symmetric 
case. This means that the energy gain 
$\chi$ (which is the selectivity parameter) for a P - monomer in its own (polar) 
environment is equal to the corresponding energy gain for a H - monomer, provided 
the latter stays in the hydrophobic environment. Also, the P - and H - parts of 
a diblock have an equal number of monomers. An estimate for the effective attraction 
energy per diblock of length $2M$ ($M$ being the number of monomers of each species
in the diblock) then reads \cite{Sommer5,Sommer6}.
\begin{eqnarray}
E \sim - \chi^2 M^2,
\label{E}
\end{eqnarray}
where we measure the energy in units of $k_B T$, (i.e. $\chi/k_BT \rightarrow \chi$ is 
dimensionless) where $k_B$ denotes the Boltzmann constant.

Now consider a multiblock copolymer of length $N$ which consists of ${\cal N}= N/2M$ 
diblocks. At the adsorption threshold, $\chi_c$, the attraction energy per chain 
scales as\cite{Eisen} $V_{\rm attr} \approx |E| {\cal N}^\phi$, where $\phi$ denotes
the so called crossover exponent. It has been shown rigorously \cite{Bray} that 
the relation $\phi = 1 - \nu$  (here $\nu = 0.6$ is the Flory exponent in $3D$ 
space) is correct for the case of a penetrable surface. 
At $\chi_c$ the attractive energy, $V_{\rm attr}=V^c_{\rm attr}$, should be of the 
order of the thermal energy, i.e. 
\begin{eqnarray}
|E^{c}| {\cal N}^{1 - \nu} \approx 1.
\label{thermal}
\end{eqnarray}
Taking into account eq. (\ref{E}), the expression for the critical selectivity 
then yields
\begin{eqnarray}
\chi_{c} \sim M^{-(1+ \nu)/2} N^{-(1 - \nu)/2}.
\label{critical}
\end{eqnarray}
Eq.(\ref{thermal}) refers to the chain as a whole at the localization threshold $\chi_c$. 
As the selectivity parameter $ \chi $ increases, each part of the chain (named "blob") 
containing $ g $ monomers accumulates an energy of the order of $ k_B T $. Note
that the chain "neutrality" requires that $g/M$ is always an even number.  
The blob length $ g $ may be determined from $ |E| (g/M)^{1 - \nu} \approx 1$, so that 
\begin{eqnarray}
g \sim \chi^{- 2/(1 - \nu)} M^{- (1 + \nu)/(1 - \nu)}.
\label{blob}
\end{eqnarray}
The localized chain as a whole can be considered as a string of such blobs.

It is customary\cite{Gennes1} to take the number of blobs, $ N/g \sim \chi^{2/(1 - 
\nu)} M^{(1 + \nu)/(1 - \nu)} N $ as a natural scaling variable in the various scaling 
relations. For the sake of convenience, we take this variable in the following form
\begin{eqnarray}
\eta \equiv \chi \: M^{(1 + \nu)/2} \: N^{(1 - \nu)/2}.
\label{argument}
\end{eqnarray}
Then the chain size perpendicular to the interface direction scales as
\begin{eqnarray}
R_{g\perp} = l N^{\nu}{\cal G}(\eta) \quad,
\label{perp}
\end{eqnarray}
where $ l $ is the Kuhn segment and ${\cal G}(\eta) $ is some scaling function. 

One should discriminate between the cases of weak and strong localization \cite{Leclerc}. 
In the weak localization regime the scaling argument $ \eta $ is of the order of one 
(cf. eq.(\ref{critical})), and the scaling function ${\cal G} (\eta) $ has a power law behavior, i.e.
\begin{equation}
{\cal G}(\eta) = 
\left\{\begin{array}{l@{\quad,\quad}l}
1 &\mbox{at} \quad \eta < 1\quad \\ \eta^{\alpha} &\mbox{at} 
 \quad \eta \geq 1
\end{array}\right. \quad,
\label{g_function}
\end{equation}  
where $ \alpha = - 2 \nu/(1 - \nu) $. This value of the exponent $ \alpha $ has been 
found from the condition that at $ \eta \geq 1 $ the value of $ R_{g\perp} $ is determined 
by the blob size alone, i.e. it does not depend on $ N $.

Similar reasoning can be used for the chain size in direction parallel to the interface 
$ R_{||} $, namely
\begin{eqnarray}
R_{g\parallel} = l N^{\nu} {\cal H}(\eta)
\label{parall} \quad,
\end{eqnarray}
where $ {\cal H}(\eta) $ is some scaling function. Again, the form of $ {\cal H}(\eta) $ 
can be determined by the condition that at $ \eta \geq 1 $ the localized multiblock chain 
behaves as a two - dimensional self - avoiding string of blobs, i.e. $ R_{||} \sim 
N^{\nu_2} $, where $ \nu_2 = 0.75 $ is the Flory exponent in $ 2 d $ space. As a result 
one finds
\begin{equation}
{\cal H}(\eta) = 
\left\{\begin{array}{l@{\quad,\quad}l}
1 &\mbox{at} \quad \eta < 1\quad \\ \eta^{\beta} &\mbox{at} 
 \quad \eta \geq 1
\end{array}\right. \quad,
\label{h_function}
\end{equation}  
where $ \beta = 2 (\nu_2 - \nu)/(1 - \nu) $.

As the selectivity parameter $ \chi $ grows further, a characteristic point 
$ \chi = \chi_{\infty} $ can be reached where $ R_{g\perp} $ and $ R_{g\parallel} $ approach 
a plateau and do not change any further. This is the strong localization limit where the 
number of monomers in the blob becomes of the order of the block length, i.e. $ g \approx M $. 
By making use of 
this condition in eq. (\ref{blob}) we can write for the characteristic selectivity  
$ \chi_{\infty} $  the following simple relation
\begin{equation}
\chi_{\infty} \sim M^{-1}.
\label{infty}
\end{equation} 
which corresponds to the well known Flory-Huggins limit of phase separation.
Note that $\chi_{\infty}$ depends only on the block size $M$ and does not depend on
the chain length $N$.
Given this value one can go back to eqs.(\ref{perp} - \ref{h_function}) and check the 
scaling in the strong localization limit. The resulting relations are
\begin{eqnarray}
R_{g\perp} &=& l M^{\nu},\label{strong1}\\
R_{g\parallel} &=&  l M^{\nu} \left( \frac{N}{M}\right)^{\nu_2} \sim 
M^{- (\nu_2 - \nu)} N^{\nu_2}.
\label{strong2}
\end{eqnarray}
which reflect the pancake geometry of a polymer localized at the surface.
The relations (\ref{strong1})  and (\ref{strong2})  can be expected because in this regime 
all P - and H - segments are predominantly in their preferred solvents. In addition,
the scaling estimate of the polymer size perpendicular to the interface shows that the 
perpendicular extension of the chain is entirely determined by the size of the blob. 

The free energy gain in the localized state is proportional to the number of blobs\cite{Gennes1}, 
i.e.
\begin{eqnarray}
F_{\rm loc} \sim \frac{N}{g} \sim \chi^{2/(1 - \nu)} N M^{(1 + \nu)/(1 - \nu)} \quad,
\label{F_loc}
\end{eqnarray}
where eq.(\ref{blob}) has been used. In terms of $\eta$ eq.(\ref{F_loc}) reads
\begin{equation}
 F_{\rm loc} = \eta^{2/(1-\nu)} \quad \mbox{at} 
 \quad 1 \leqslant \eta  \leqslant \eta_{\infty},
\label{f_function}
\end{equation} 
where $ \eta_{\infty} = \chi_{\infty} M^{(1+\nu)/2}N^{(1-\nu)/2} \sim (N/M)^{(1-\nu)/2}$.
At $ \chi  >  \chi_{\infty}$ the free energy gain follows the strong localization law
\begin{equation}
F_{\rm loc} \sim \chi N \quad,
\label{F_strong}
\end{equation}
which is mainly triggered by the energy gain.

\subsection{Dynamics}

After the localization on the selective interface, the multiblock
copolymer can only diffuse in the two - dimensional space along the
surface. Here we give the scaling estimate for the characteristic
time $\tau $ which is necessary for a chain to displace along the
surface a distance equal to its own gyration radius. We define 
this time in terms of the chain length $N$ and the  block length
$M$ in the strong localization limit.

The diffusion coefficient of the whole chain along the surface can be
estimated as 
\begin{eqnarray}
D_{\parallel} \approx \left(\frac{1}{\zeta}\right) \frac{1}{{\cal N}} =
\frac{D_{\rm block}}{{\cal N}},
\label{Diff_coef}
\end{eqnarray}
where $D_{\rm block}$ is the diffusion coefficient  of one block,
$\zeta = \zeta_0 M$ the corresponding friction coefficient and the
number of diblocks ${\cal N} = N/2M$. The characteristic time of the $2$
- dimensional  chain displacement obeys
\begin{eqnarray}
\tau \approx \frac{R_{\rm block}^2 {\cal N}^{2
    \nu_2}}{\left(\frac{1}{\zeta}\right)\frac{1}{{\cal N}}}, 
\label{Time}
\end{eqnarray}
where the characteristic size of the block (which spreads in the
$3$ - dimensional space) $R_{\rm block} \approx l M^{\nu}$ ($\nu = 
0.6$) whereas the $2$ - dimensional Flory exponent is $\nu_2 = 0.75$.

Now one can express all quantities in eq.(\ref{Time}) in terms of $N$ and
$M$. As a result one obtains
\begin{eqnarray}
\tau &\approx&  \zeta_0 \: l^2 M^{2 (\nu - \nu_2)} N^{2
    \nu_2 + 1}\nonumber\\
 &\approx& \zeta_0 \: l^2  M^{- 0.3} \: N^{2.5}.
\label{Time_final}
\end{eqnarray}
It is clear that this result should hold for sufficiently long chains 
and blocks.

\section{MC - simulation}

\subsection{Description of the model}

The off-lattice bead-spring model has been used previously for simulations of polymers 
both in the bulk \cite{36,37} and near confining surfaces \cite{33,35,38,39,40,41}; thus, 
we describe here the salient features only. Each polymer chain contains $N$ effective 
monomers connected by anharmonic springs described by the finitely extendible nonlinear 
elastic (FENE) potential.

\begin{equation} \label{eq1}
U_{FENE} =-\frac{K}{2} R^2 \ln \big[1-\frac{(\ell - \ell_0)^2}{R^2} \big].
\end{equation}
Here $\ell$ is the length of an effective bond, which can vary in between $\ell_{min}< \ell
<\ell_{max}$, with $\ell_{min}=0.4$, $l_{max}=1$ being the unit of length, and has the
equilibrium value $l_0=0.7$, while $R=\ell_{max}-\ell_0=\ell_0-\ell_{min}=0.3$, and the spring
constant $K$ is taken as $K/k_BT=40$. The nonbonded interactions between the effective 
monomers are described by the Morse potential

\begin{equation} \label{eq2}
U_M=\epsilon_M \{\exp[-2 \alpha(r-r_{min})]-2 \exp [-\alpha(r-r_{min})]\} \, ,
\end{equation}
where $r$ is the distance between the beads, and the parameters are chosen as $r_{min}=0.8$, 
$\epsilon_M=1$, and $\alpha=24$. Owing to the large value of the latter constant, $U_M(r)$ 
decays to zero very rapidly for $r>r_{min}$, and is completely negligible for distances larger 
than unity. This choice of parameters is useful from a computational point of view, since it 
allows the use of a very efficient link-cell algorithm \cite{42}.  From a physical point
of view, these potentials 
eqs.~(\ref{eq1}),~(\ref{eq2}) make sense when one interprets the effective bonds as
kind of Kuhn segments, comprising a number of chemical monomers along the chain, and thus the 
length unit $\ell_{max}=1$ corresponds physically rather to 1 nm than to the length of a 
covalent $C-C$ bond (which would only be about $1.5 \text{\r{A}}$). Since in the present study 
we are concerned with the localization of a copolymer at good solvent conditions, in 
eq.(\ref{eq2}) we retain the repulsive branch of the Morse potential only by setting
$U_M(r) = 0 \quad \mbox{for} \quad r > r_{min}$ and shifting $U_M(r)$ up by $\epsilon_M$.

The interface potential is taken simply as a step function with amplitude $\chi$,
\begin{eqnarray}
\label{eq3}
U_{int}(n,z)=
\begin{cases}-\sigma(n)\chi/2, &z > 0  \\
\sigma(n)\chi/2, &z \le 0 
\end{cases}
\end{eqnarray}
where the interface plane is fixed at $z = 0$, and $\sigma(n) = \pm 1$
denotes a "spin" variable which distinguishes between P- and H- monomers. The energy
gain of each chain segment is thus $-\chi$, provided it stays in its preferred solvent.

\begin{figure}
\noindent
\begin{minipage}[t]{.49\linewidth}
\epsfig{file=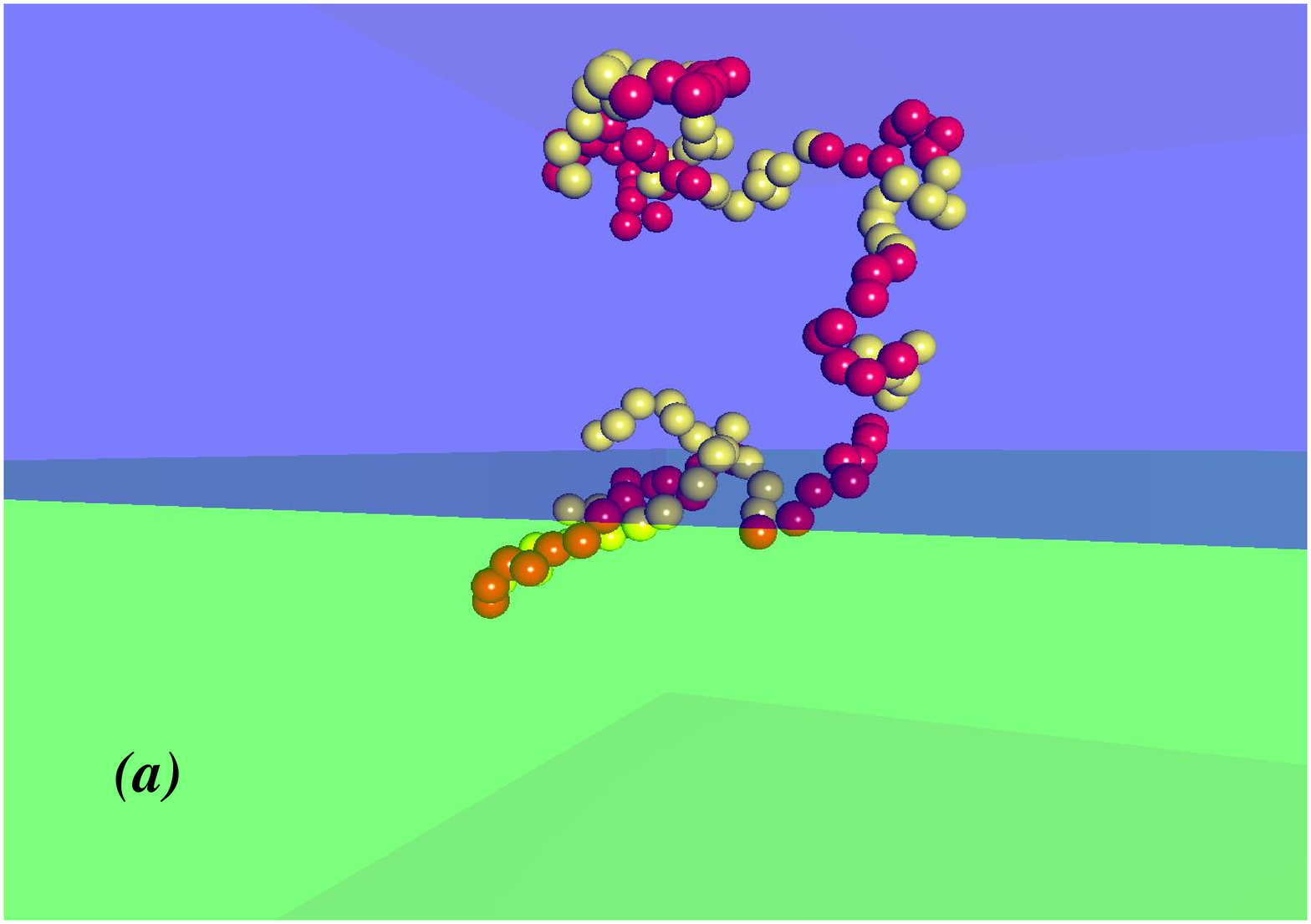,width=1.0\linewidth} 
\end{minipage}\hfill
\begin{minipage}[b]{.46\linewidth}
\epsfig{file=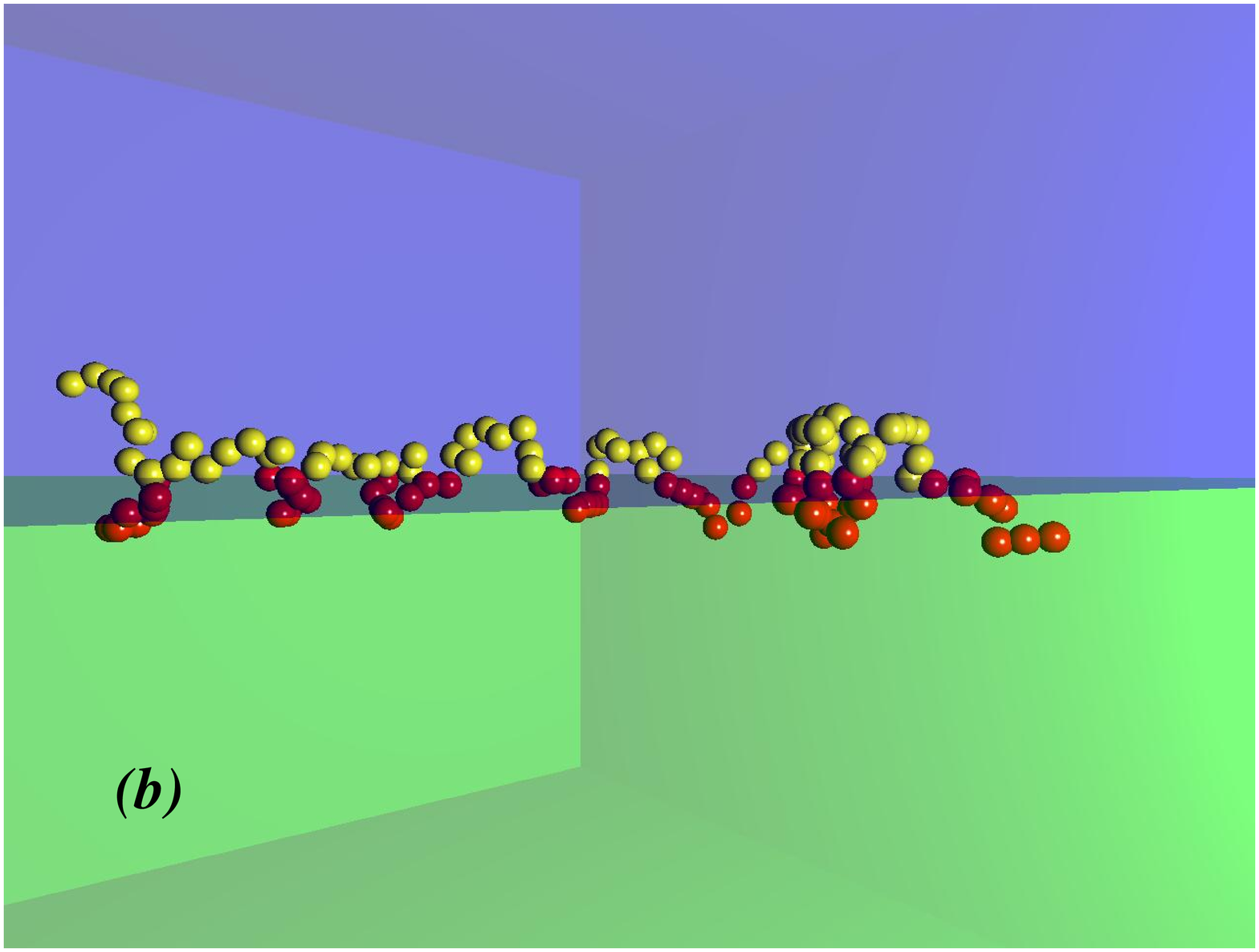,width=1.0\linewidth}
\end{minipage}
\caption{Snapshots of typical configurations of a copolymer with $N = 128$, $M = 8$ at 
$ \chi = 0.25 $ (a), and $ \chi = 10$ (b). 
The value of the critical selectivity for this 
chain is $ \chi_c = 0.67 $.}
\label{Snapshot}
\vspace{1cm}
\end{figure}

Two typical snapshots for a chain with $ N = 128 $ and $ M = 8 $ in the weak 
($ \chi = 0.25 $) and strong ($ \chi = 10 $) localization limits are shown in 
Fig.\ref{Snapshot} to illustrate the main features of the model.

In each Monte Carlo update, a monomer is chosen at random and one attempts to displace 
it randomly by displacements $\Delta x, \: \Delta y,\: \Delta z$ chosen uniformly from 
the intervals $-0.5\le \Delta x, \: \Delta y,\: \Delta z\le 0.5$. The transition
probability for such an attempted move is simply calculated from the total change $\Delta 
E$ of the potential energies defined in eqs. (\ref{eq1} - \ref{eq3}) as $W=\exp(-\Delta E/
k_BT)$. According to the standard Metropolis algorithm, the attempted move is accepted only
if $W$ exceeds a random number uniformly distributed between zero and unity. Since our
potentials are constructed such that chains cannot intersect themselves in the course of
random displacement of beads, one does need to check separately for entanglement restrictions.
In the course of the simulation the starting configuration of the copolymer is relaxed
for a period of time $\tau_{0}$ before measurements of the chain properties are carried out.
We sample various static and dynamic quantities as the components of the gyration radius
perpendicular, $R_{g\perp}$, and parallel, $R_{g\parallel}$, to the interface, the
density distribution of the two kinds of monomers around the interface, internal energy,
specific heat, diffusion coefficients, etc. We use periodic boundary conditions in the plane
of the interface while there are rigid walls in the z - direction, where the simulation box
extends from $z = -16$ to $z = 16$.
The algorithm is reasonably fast, i.e. one
performs $\approx 0.5\times 10^6$ updates per CPU second on a $2.8$ GHz PC.
Typically we studied chains with length $32 \le N \le 512$ and all measurements have been
averaged over $2^{17}$ samples.

\subsection{Results}

\begin{figure}[ht]
\begin{center}
\epsfig{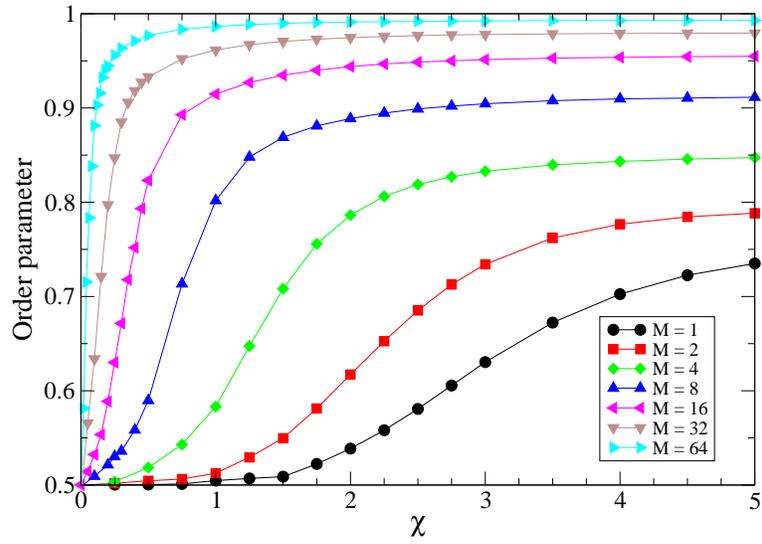} 
\caption{The order parameter for chain length $ N=128 $ vs selectivity $\chi$ 
for different values of the block length $M$.}
\label{OP}
\end{center}
\end{figure}

The localization at the selective interface can be considered as a phase transition at 
least in the limits: $ N\rightarrow \infty $, $ M\rightarrow \infty $ and $ \chi 
\rightarrow 0 $ with $ \eta_c \equiv  \chi_c N^{(1-\nu)/2} M^{(1+\nu)/2} = 1$. 
At finite $ N,M $ and $ \chi $ the transition looks like a smooth crossover and one can 
define an order parameter (OP) in terms of the fractions of P - and H - 
monomers in the polar (at $z > 0$) and hydrophobic (at $z < 0$) semispace, respectively, i.e.
\begin{equation}
\mathrm{OP} = \sqrt{f_{P}^{>}f_{H}^{<}}
\end{equation}
Fig.\ref{OP} shows the variation of this OP  with the selectivity parameter $\chi$ for a 
chain length $N = 128$ and various lengths of the blocks $M$. Evidently, for 
sufficiently large $M$ the OP reaches saturation at much lower values of $\chi$ than for 
short blocks. It can also be seen that the chains with small $M$ even in the localized 
state display a much lower degree of ordering than those with longer blocks since 
in the localized phase the number of "frustrated" monomers which stay in the 
wrong solvent increases with vanishing proximity to the interface.

\begin{figure}[ht]
\begin{center}
\epsfig{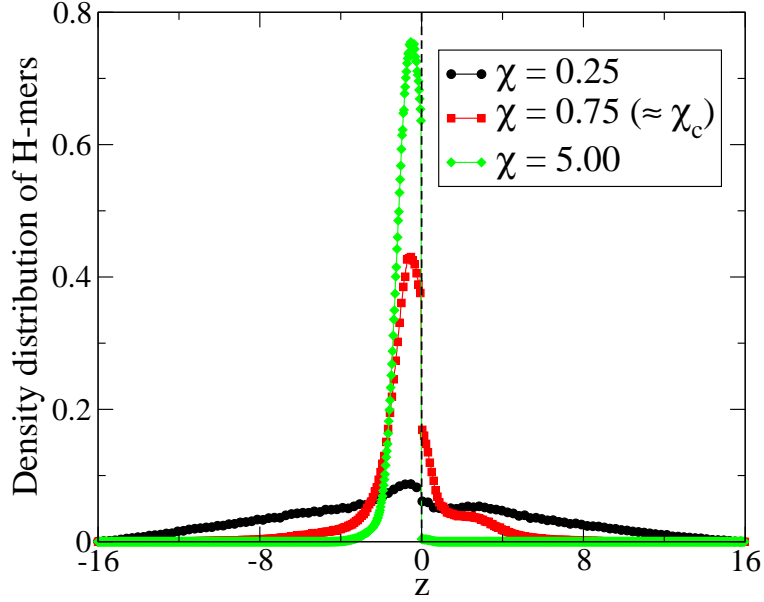} 
\caption{Density distribution of the $H$ - monomers for a chain with $N = 128$ and 
$M = 8$ at  three different values of $\chi$. The critical selectivity is $\chi_c
\approx 0.67$.}
\label{Histo}
\end{center}
\end{figure}

The transition location, $\chi_c$, defined as the inflection  point of the curves 
in Fig. \ref{OP} can be easily calculated. A more comprehensive information about 
the localization is revealed from the monomer density distribution histogram which 
is given in Fig. \ref{Histo}.
While for $\chi=0.25$ the copolymer is still not localized, even at $\chi =0.75 >
\chi_c$ there is still a considerable amount of segments which stay in the wrong
solvent and one needs a rather strong segregation $\chi = 5.0$ in order to keep
the segments entirely in their preferred environment.

\begin{figure}[ht]
\epsfig{file=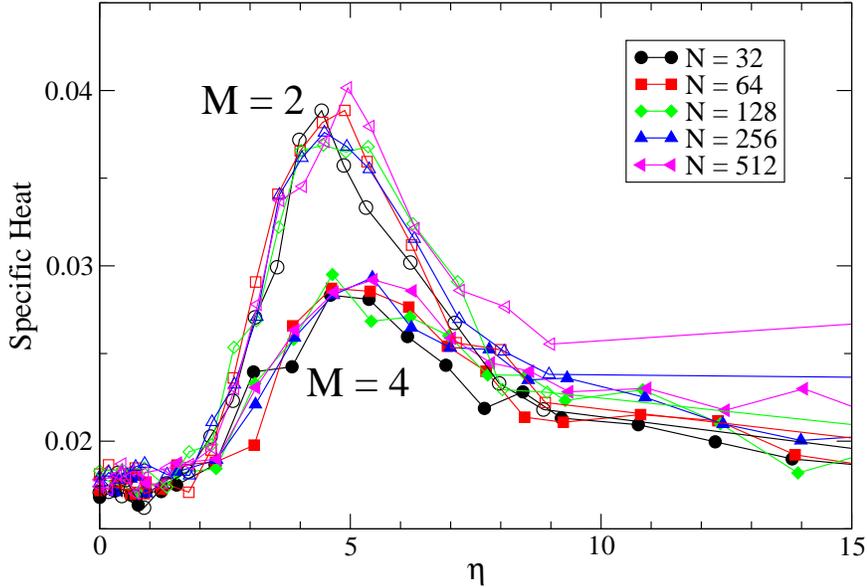,width=10cm,angle=270}
\caption{Specific heat (per monomer)of a copolymer chain with length $32 \le N\le 512$
and two block sizes $M=2, 4$ at the localization transition.}
\label{Cv}
\end{figure}

The localization at the interface is also marked by a maximum in the specific heat
of the system - $C_v(\chi)=(Nk_bT^2)^{-1} \left(\langle E^2\rangle - \langle E\rangle
^2\right)$ - which we obtain from the fluctuations of the internal energy of the
copolymer $E$ and show in Fig. \ref{Cv}.
One can readily see from Fig. \ref{Cv} that the specific heat curves collapse (despite 
significantly larger scatter in the measured data) onto different master curves 
for the two block sizes in concern. However, the heights of the $C_v$-maxima at the 
localization transition appear rather insensitive with respect to chain length $N$. 
Indeed, it has been shown earlier\cite{EisenBook} that the specific heat critical 
exponent is zero, and there exist only weak logarithmic corrections to finite size
scaling ($C_v \sim (ln N)^{3/11}$ for a Gaussian chain). One can also infer from
Fig. \ref{Cv} a rapid decrease in the height of the $C_v$-maximum with growing block
size $M$ which is to be expected bearing in mind that the chance for an energy fluctuation
rapidly diminish as less and less monomers cross into the "wrong" solvent with
increasing $M$.

\subsubsection{Weak localization}

\begin{figure}[htb]
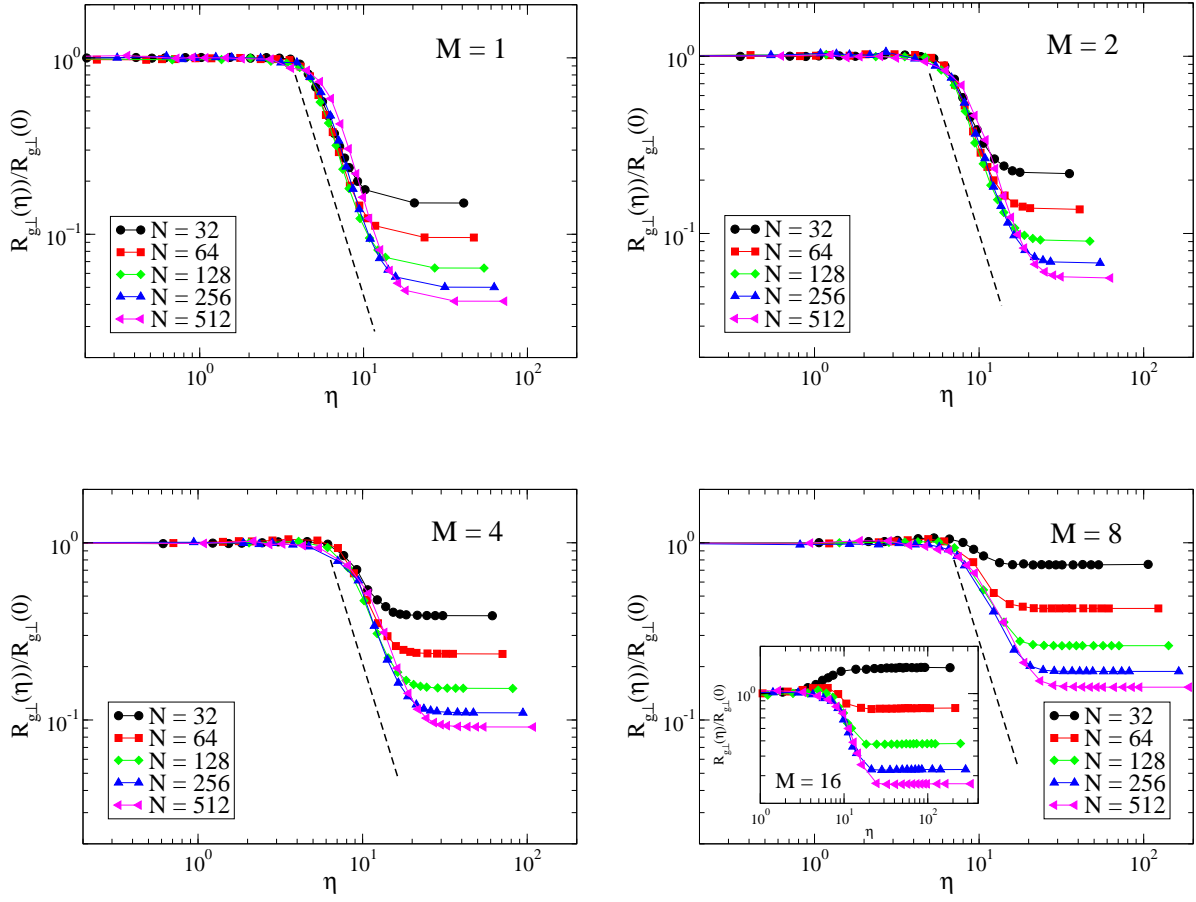

\noindent
\begin{minipage}[b]{.46\linewidth}
\epsfig{file=sommer_scaling_M1.eps,width=1.0\linewidth}
\end{minipage}\hspace{0.5cm}
\begin{minipage}[b]{.46\linewidth}
\epsfig{file=sommer_scaling_M2.eps,width=1.0\linewidth}
\end{minipage}
 
\vspace{1cm}
\begin{minipage}[b]{.46\linewidth}
\epsfig{file=sommer_scaling_M4.eps,width=1.0\linewidth}
\end{minipage}\hspace{0.5cm}
\begin{minipage}[b]{.46\linewidth}
\epsfig{file=sommer_scaling_M8.eps,width=1.0\linewidth}
\end{minipage}
\caption{Variation of the chain size perpendicular to the interface $ R_{g\perp} $ 
with the scaling variable $ \eta $ for block lengths $M=1, 2, 4, 8\: \mbox{and}\: 16 
\: \mbox{(inset)}$. In all cases the dashed line indicates the scaling prediction.}
\label{Sommer_scaling}
\end{figure}

We turn now to the comparison of our data with the scaling prediction given in Sec.II.  
Figure \ref{Sommer_scaling} displays the variation of the chain size perpendicular to 
the interface
$ R_{g\perp} $ with scaling parameter $ \eta $. One can easily distinguish between the
cases of weak and strong localization. In the first case the curves collapse nicely on a 
master curve in a range of $ \eta $ which gets narrower with growing $M$. This proves 
the scaling law given by eq. (\ref{g_function}) and is also in line with the trend given 
by eq. (\ref{infty}) for the characteristic selectivity $ \chi_{\infty} $ at the onset of
the strong localization regime. The scaling function, eq. (\ref{g_function}), is characterized 
by the exponent $ \alpha = -3 $, which is indicated by the dashed line in Fig. 
\ref{Sommer_scaling} and appears in good agreement with the simulation data. An 
interesting case which we observe in all our studies is shown in the inset of Fig. 
\ref{Sommer_scaling} - in the case of a {\em diblock} (i.e. when $M=N/2$) in the strong 
localization regime one finds $R_{g\perp}(\eta)/R_{g\perp}(0) \ge 1$. Clearly, as
the two blocks tend to stay away from the interface, thus maximizing entropy, they
induce an elongation of the coil perpendicular to the interface in contrast to the 
behavior of multiblock copolymers.

\begin{figure}[ht]
\epsfig{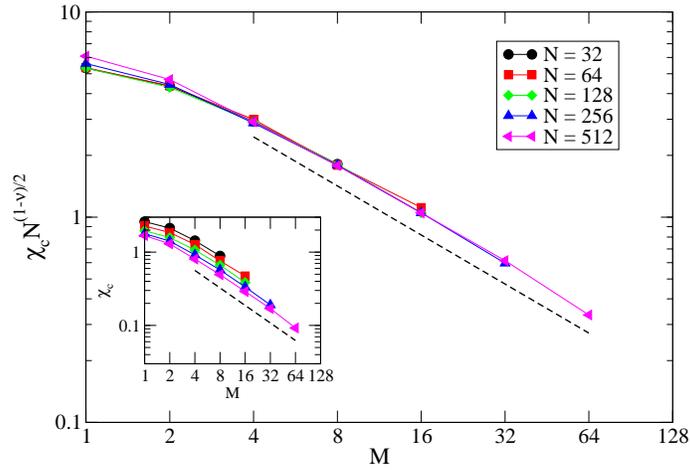} 
\caption{Scaling plot of the critical selectivity $\chi_c$ in terms of block length $M$ 
and chain length $N$. The dashed line indicates the scaling prediction. The inset 
shows $\chi_c$ vs $M$ without scaling in $N$.}
\label{Chi_c}
\end{figure}

The set of graphs shown in Fig. \ref{Sommer_scaling} has been used to evaluate the 
critical $ \chi_c $ taken as the positions of the respective inflection points. 
The result of this evaluation 
is shown in Fig. \ref{Chi_c}. The insert shows the $M$ - dependence of $\chi_c$ at 
different $N$.
Evidently, the rescaling according to eq. (\ref{critical}) leads to the 
master curve. The dashed line demonstrates the scaling prediction $ \chi_c \sim M^{-0.8} $ 
and  is again in a good agreement with theory. Not surprisingly, at small $M$ one
observes deviations from the expected power-law behavior which demonstrate that 
scaling validity is achieved in the asymptotic limit only.

\begin{figure}[htb]
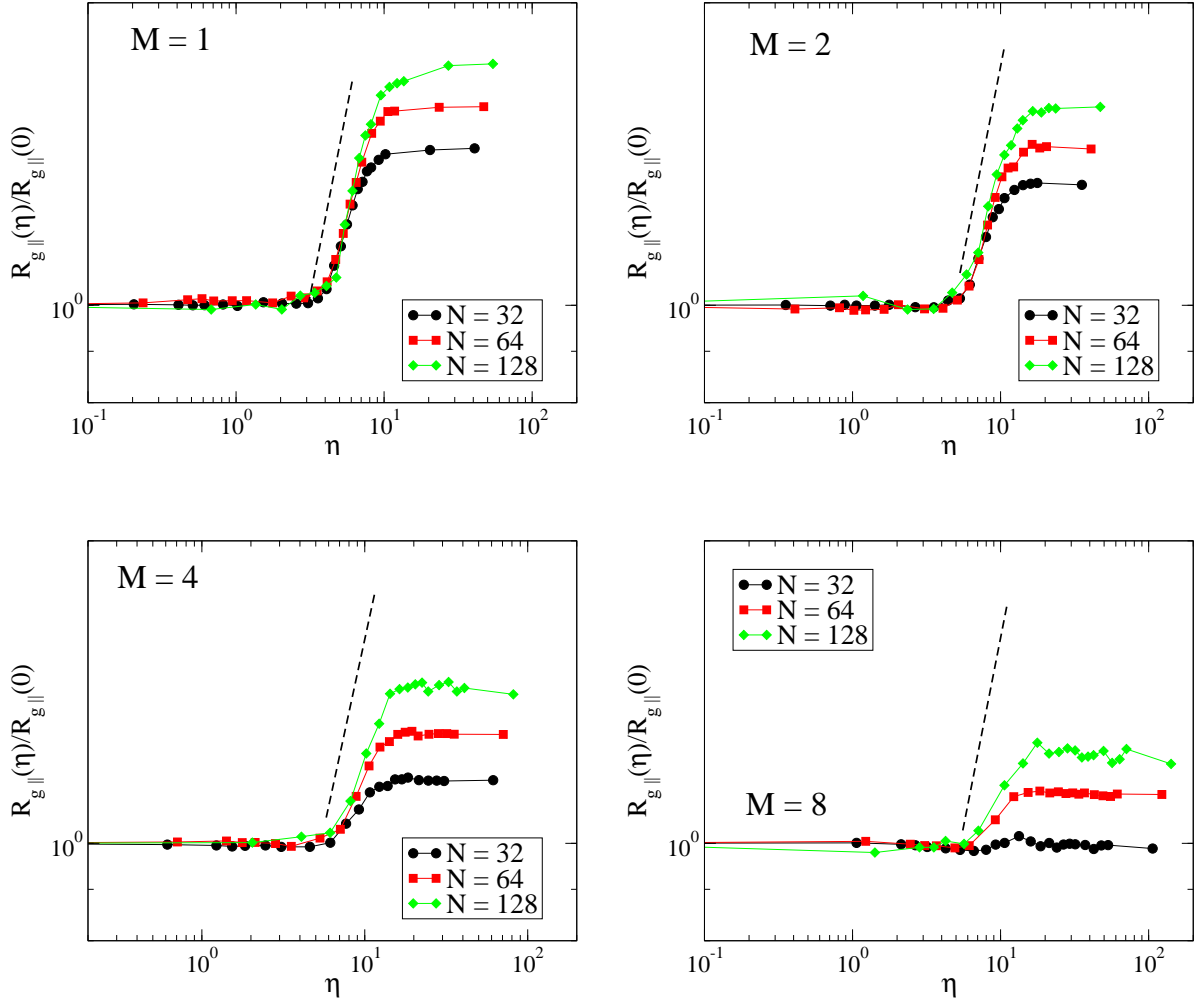

\noindent
\begin{minipage}[b]{.46\linewidth}
\epsfig{file=sommer_scaling_M1_parall.eps,width=1.0\linewidth}
\end{minipage}\hspace{0.5cm}
\begin{minipage}[b]{.46\linewidth}
\epsfig{file=sommer_scaling_M2_parall.eps,width=1.0\linewidth}
\end{minipage}

\vspace{1cm}

\begin{minipage}[b]{.46\linewidth}
\epsfig{file=sommer_scaling_M4_parall.eps,width=1.0\linewidth}
\end{minipage}\hspace{0.5cm}
\begin{minipage}[b]{.46\linewidth}
\epsfig{file=sommer_scaling_M8_parall.eps,width=1.0\linewidth}
\end{minipage}
\caption{Variation of $ R_{g\parallel} $ with the scaling variable $ \eta $ for
chain lengths $32 \le N \le 512$ and several block lengths, as indicated. The
onset of the strong localization regime is manifested by the decay of the master
curves into different curves for each $N$. In all cases the dashed line indicates
the scaling prediction.}
\label{Scaling_paral}
\end{figure}

Similar scaling representation has been done also for the parallel component of the chain size 
$ R_{g\parallel} $ (see Fig. \ref{Scaling_paral}). In the weak localization regime all data 
collapse again into a master curve. The scaling exponent $\beta = 3/4$ (noted by the
dashed line, see eq.(\ref{h_function}) is fairly close to the simulational results.

\subsubsection{Strong localization}

\begin{figure}[hb]
\epsfig{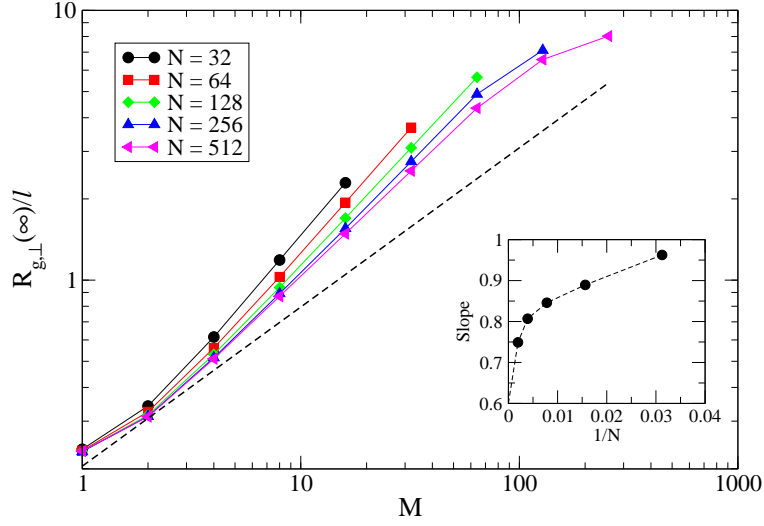} 
\caption{The perpendicular component $ R_{g\perp}(\infty)/l $ as a function of $M$ at the 
different $N$. The dashed line corresponds to the scaling prediction. The insert shows the 
extrapolation for $ 1/N \rightarrow 0 $ as leading to the  theoretical value of the slope 
$\nu = 0.6$. }
\label{Plateau}
\end{figure}

\begin{figure}
\epsfig{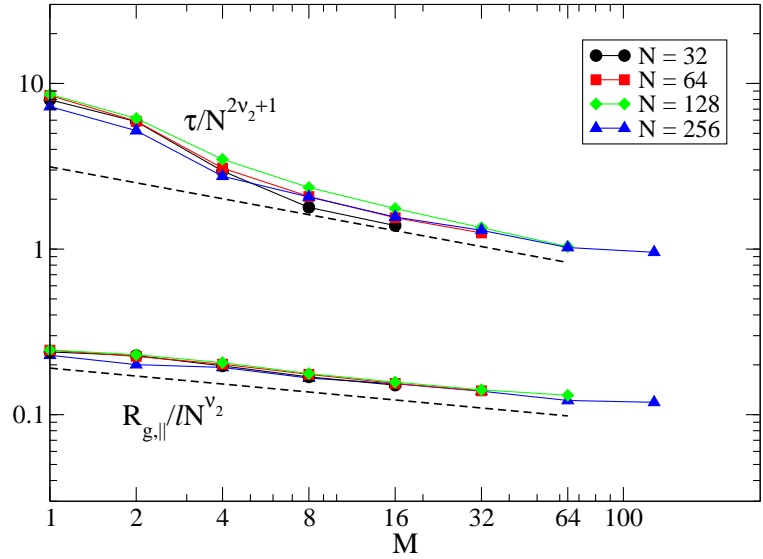} 
\caption{The upper curve represents the rescaling of the characteristic time $\tau$ 
according to eq. (\ref{Time}). The lower curve shows that the scaling for 
$ R_{g\parallel}(\infty) $ follows eq.(\ref{strong2}). The dashed lines indicate the 
scaling predictions.}
\label{Dynamics}
\end{figure}

The area where $ R_{g\perp} $ and $ R_{g\parallel} $ abandon the master curves in Figs. 
\ref{Sommer_scaling}, \ref{Scaling_paral} and approach different plateau values at high 
$\eta$ corresponds to the strong localization regime. In this case the scaling should 
follow eqs. (\ref{strong1}) and (\ref{strong2}). We have evaluated $ R_{g\perp} (\infty)$ 
from the plateau height values in Fig. \ref{Sommer_scaling} (dropping out the factor 
$ 1/R_{g\perp}(0) $). The corresponding results are shown in Fig. \ref{Plateau}. The $M$ 
- dependence of $ R_{g\perp}(\infty)/l $ is given for different chain lengths $N$. The 
finite size analysis indicates that an extrapolation to $ 1/N \rightarrow 0 $ leads 
to the correct Flory scaling $ R_{g\perp}(\infty) \sim M^{0.6} $ (see the dashed line in 
Fig. \ref{Plateau}). 
Fig. \ref{Dynamics} makes it also apparent that the parallel component 
$ R_{g\parallel}(\infty) $ follows to a good accuracy the law $ R_{g\parallel}(\infty)/l 
\sim M^{-0.15} N^{0.75} $ which comes from eq. (\ref{strong2}). Thus Fig. \ref{Dynamics} 
demonstrates as well the scaling relation for the relaxation time $\tau$ given by 
eq.(\ref{Time}). The fit of our MC - results with the theoretical scaling prediction 
$ \tau \sim M^{-0.3} N^{2.5} $ is very good especially taking into account the relatively 
small values of the block length.

\section{Conclusion and outlook}

In the preset work we have studied the adsorption of regular HP copolymer at a penetrable
interface between two immiscible fluids. Specifically, we have focused our 
attention on the influence of the block size $M$ on the adsoprtion transition as it is
known that efficient compatibilizing agents (i) reduce the surface tension and (ii)
enhance the adhesion between the surfaces that result upon cooling. While reduced
surface tension is achieved by possibly higher surface coverage by the polymer, for
enhanced adhesion between the components each block should extend and penetrate a
significant distance into the respective compatible melt.  Qualitatively it is clear
that shorter blocks will provide a more efficient reduction of the surface tension 
whereas longer blocks would favor better adhesion between the segregated systems.
In order to get more quantitative insight into the effectiveness of multiblock copolymers  
we develop a simple scaling treatment which is then compared against simulational results. 

Within the framework of our simple scaling approach we arrive at several important 
conclusions characterizing the impact of block length $M$ on the behavior of regular 
multiblock copolymers at a fluid-fluid interface:
\begin{itemize}
\item The critical selectivity decreases with growing block length as $\chi_c \sim
M^{-(1+\nu)/2}$ while the crossover selectivity to the strong localization 
regime vanishes as $\chi_{\infty} \sim M^{-1}$.
\item The size of the copolymer varies in the weak localization regime as
$R_{g\perp} \sim M^{-\nu(1+\nu)/(1-\nu)}$ and $R_{g\parallel} \sim M^{(\nu_2 - \nu)
(1+\nu)/(1-\nu)}$.
\item In the case of strong segregation we obtain respectively $R_{g\perp}\sim M^\nu$
and $R_{g\parallel} \sim M^{-(\nu_2-\nu)}$.
\item The typical relaxation time in the case of strong localization varies as $\tau \sim
M^{2(\nu-\nu_2)}$ with block length.
\end{itemize}
Our computer experiments appear to confirm nicely these predictions.

Concluding, one should point out that this study has been focused on one aspect of
the localization of copolymers at penetrable interfaces. A number of related topics
such as the case of many chains in the system, the presence of attractive interactions
between similar segments, or the kinetics of localization could provide further
valuable insights on the problem. Another interesting aspect comes when random - or 
specifically sequenced copolymers are considered rather than regular multiblock copolymers. In 
these cases additional information is stored in the sequence of the H and P monomers. The
relevant information can be read by the interfacial behavior of such polymers. Indeed, the 
localization will then depend on the distribution of the H and P monomers along the chain. 
For special sequences we expect to relate their behavior to some typical proteins \cite{lee}. 
We hope to be able to report on some of these topics in the near future.

\section{Acknowledgments}

AM acknowledges the support and hospitality of the Max-Planck Institute for
Polymer Research in Mainz during this study. This research has been supported by the 
Sonderforschungsbereich (SFB 625).

\clearpage

\end{document}